\documentclass[]{spie}  

\usepackage{amsmath,amsfonts,amssymb,float}
\usepackage{graphicx}
\usepackage[colorlinks=true, allcolors=blue]{hyperref}
\usepackage{tabularx, booktabs, makecell, caption} 
\usepackage{siunitx}
\usepackage[raggedrightboxes]{ragged2e}

\title{Centroiding Undersampled PSFs with a Lookup Table}

\author[a]{Kevin J. Ludwick}
\author[b]{Ashley Mazingo}
\affil[a]{University of Alabama in Huntsville, 301 Sparkman Dr, Huntsville, AL 35899, USA}

\authorinfo{Further author information: (Send correspondence to K.J.L.)\\K.J.L.: E-mail: kevin.ludwick@uah.edu, Telephone: 1 256 824 2527}

\usepackage{lineno}

\begin{document}
\maketitle

\begin{abstract}
We present a method of centroiding undersampled point spread functions (PSFs) that may be useful, especially when dithering is not an option.  If the profile of the expected PSF is known fairly well through characterization of the telescope and detector used for observing, one can simulate the undersampled PSF at many positions on a simulated pixel grid.  The true centroid positions are known since the PSFs are simulated, and so one can match up each undersampled PSF images to its true centroid location, thus forming a lookup table.  One then assigns the centroid position of an observed PSF to the position associated with the PSF in the lookup table that has the smallest squared residual with respect to the observed PSF.  We examine a few PSF sizes and demonstrate that the lookup table provides better centroid positions compared to a fitting algorithm when the PSFs are undersampled, even in the presence of noise.
\end{abstract}


\section{INTRODUCTION}
\label{sec:intro}  
Reliable centroiding of point spread functions (PSFs) in a star scene is often crucially important for astronomical observations.  Accurate determination of angular distances between stars and pointing directions relies on accurate and consistent centroiding, but this is in general not possible with a simple fit when PSFs are undersampled (sampled at a rate less than what is needed for Nyquist sampling).  When prominent features of a PSF lie on non-integer pixel values, the undersampling of the pixels does not faithfully represent the PSF, and consistent centroiding is not achievable with a simple fit.  

One method of achieving better resolution of the PSF is through dithering \cite{PASP, Anderson2, Zhuang_2024}.  However, there may be cases when dithering is not desired or not possible (for fast tracking of particular targets, when observation and processing time is short, when observations are done in parallel \cite{Rigby_2023}, or when the physical constraints of a lab testbed do not allow for dithering).  In this case, if the expected PSF can be modeled well, whether through software or through examination with an interferometer, a lookup table which pairs simulated PSFs with their known simulated centroid positions can provide a useful method of centroiding.  

In the following sections, we examine the case of Gaussian PSFs of various sizes.  We discuss our methodology of creating the lookup tables and our results when employing them in finding the centroid for various test PSFs.  We find that the lookup table more accurately centroids undersampled PSFs compared to a fitting algorithm.

\section{METHODOLOGY FOR LOOKUP TABLES}
We use a script to create a lookup table of PSF images and the associated true x, y centroid positions. 
The PSF used in this work's analysis is the that of a 2D Gaussian.  Various parameters, which exert control over PSF image dimensions, size, and pixel values, are configured. The script proceeds to generate a grid of x, y coordinates by iterating through x and y values for centroid location within an interval spanning 1 pixel. The x, y centroid locations within this interval are incremented by some step size. In this work, we use a 15x15 pixel area, a step size of 0.1 pixels, and a centroid location that would round to pixel location (8,8), so PSFs are generated with centroid locations at (7.5,7.5), (7.5,7.6), (7.5,7.7), ..., (7.5,8.5), (7.6,7.5), (7.6,7.6), (7.7,7.7), ..., (8.5, 8.5).  For each combination of $(x_0, y_0)$ centroid coordinates, we create a frame containing a noiseless Gaussian PSF. For the Gaussian, we use 
\begin{equation}
PSF(x,y) = C + A \exp\left[\frac{-(x-x_0)^2}{2 \sigma^2} -\frac{(y-y_0)^2}{2 \sigma^2}\right],
\end{equation}
where $C$ is a pedestal upon which the Gaussian of amplitude $A$ and standard deviation $\sigma$ sits.  It is centered at $(x_0, y_0)$.  We study a rotationally symmetric Gaussian here (i.e., $\sigma_x=\sigma_y\equiv \sigma$), but if we were modelling a more general PSF, we would need to include in the lookup table different orientations of the PSF for each simulated position.  For a given pixel, where the pixel location is based on the center of the physical pixel, the PSF function should be integrated over that square area centered on the pixel to obtain the intensity value assigned to it:
\begin{equation}
\int_{p_x-1/2}^{p_x+1/2} \int_{p_y-1/2}^{p_y+1/2} dx dy ~PSF(x,y),
\end{equation}
where $(p_x, p_y)$ are the integer coordinates for the physical center of the pixel.  When $\sigma$ is big enough so that the PSF is Nyquist-sampled or better, the features of the PSF are theoretically recoverable, and thus a Gaussian fitting algorithm should find the centroid well.  A good minimum PSF full-width half-maximum (FWHM) for Nyquist sampling is 2 pixels/FWHM\cite{Robertson}.  For a Guassian PSF, this means
\begin{equation}
FWHM \equiv 2 \sqrt{2 \ln 2} ~\sigma \geq 2 {\rm ~pixels},
\end{equation}
which implies 
\begin{equation}
\sigma \geq 0.85 {\rm ~ pixels}.
\end{equation}
 A dictionary is populated with the 2D array of each binned PSF as a key and the true x, y coordinates of the PSF’s centroid as the corresponding value, forming a lookup table.  The smaller the step size used to make the lookup table, the more entries there are, which allows for more accurate mapping from PSF array to centroid location.

\begin{figure}
    \centering
    \includegraphics[width=0.75\linewidth]{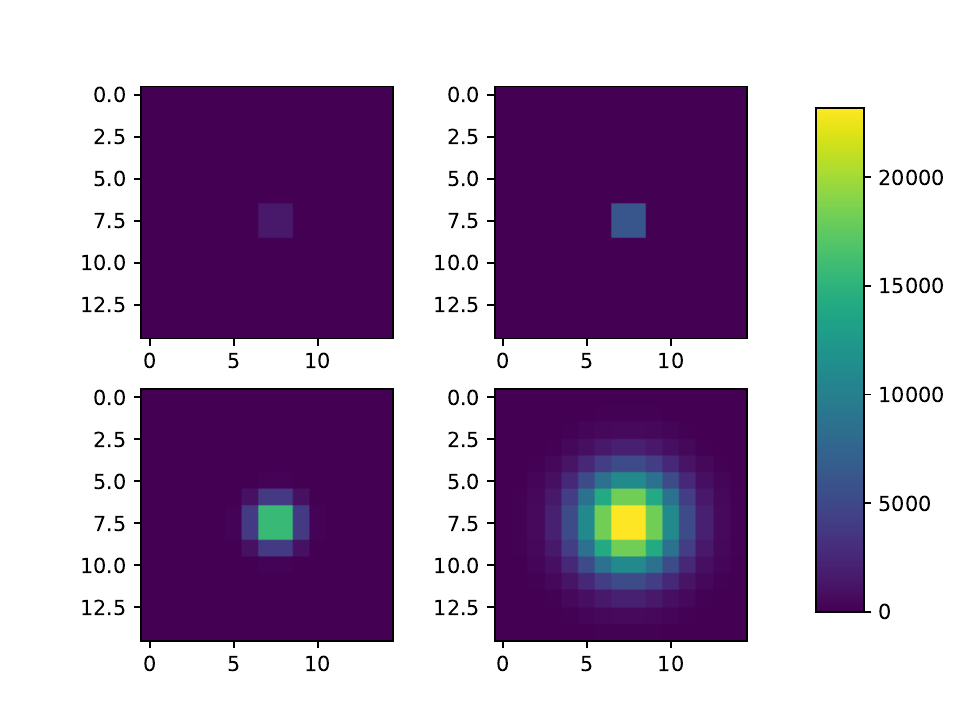}
    \caption{Gaussian PSFs. Top left:  $\sigma=0.2$ pixels. 
 Top right:  $\sigma=0.4$ pixels.  Bottom left:  $\sigma=0.8$ pixels, just below the level of Nyquist sampling ($\sigma=0.85$ pixels).  Bottom right:  $\sigma=2$ pixels, well above the level of Nyquist sampling. }
    \label{fig:enter-label}
\end{figure}

The accuracy of estimating centroid positions of PSFs using the lookup table is then evaluated. PSF images are generated using the same method used to generate PSF images for the lookup table. These test PSFs have random, uniformly distributed x, y position values within the range of positions in the lookup table. To create a realistic observed PSF, we simulate detector noise and dark subtraction.  Simulated detector noise is applied to 5 copies of each PSF frame with the publicly available module {\tt emccd\_detect}, and the 5 PSF frames are averaged. Then 5 corresponding simulated dark frames with noise are then averaged, and the averaged dark frame is subtracted from the averaged PSF frame (see Fig. \ref{fig:enter-label}). For each test PSF, centroiding is performed with both a traditional fitting algorithm and the lookup table-based method. In the lookup table-based approach, each test PSF is compared to the keys in the lookup table, and squared residuals are calculated for the PSF being evaluated with respect to each PSF stored in the lookup table. The smallest squared residual is found, and the PSF key corresponding to this residual maps to the x, y centroid position value from the lookup table. An estimated x, y position for each random PSF is then provided by the fitting algorithm method. The script then compares the true x, y positions of the tested PSFs against the positions estimated by both centroiding methods. The root-mean-squared error (RMSE) for the lookup table approach and the RMSE for the fitting algorithm approach are calculated over all test PSFs.

We examined 4 sets of PSFs, distinguished by the value of $\sigma$.  The RMSE value for each set was obtained by testing 900 random 
Gaussian PSFs with centroid locations that were in general different from the locations used to create the lookup tables. They also had randomized amplitudes (i.e., different in general from the ones used to create the lookup tables).  Normalization of test PSFs with respect to lookup table entries was applied to make the test PSFs comparable to PSFs in the lookup table.  

The uncertainty of the RMSE was calculated using\cite{RMSE} 
\begin{equation}
\frac{RMSE}{\sqrt{2 N}},
\end{equation}
where $N=900$, the number of PSFs tested for a given set.

We expected the lookup table method to result in an RMSE value lower than what the fitting algorithm method produced for $\sigma < 0.85$, assuming a sufficiently fine step size of 0.1 pixels for the creation of the lookup table.  The results of the analysis are summarized in Table \ref{noise_table}.  

\section{RESULTS}

In Table \ref{noise_table}, for the case where detector noise is included (top portion of table), the lookup table method has the lower RMSE value for all values of $\sigma < 0.85$ (i.e., a sampling rate less than Nyquist sampling).  The fitting algorithm ``wins" (has the lower RMSE value) for high sampling, $\sigma=2$ pixels, and this is expected for a sufficiently coarse step size for the lookup table.  A smaller step size would have yielded a lower RMSE value for the lookup table, at the expense of a larger lookup table, but the fitting algorithm is expected to win for well-sampled PSFs in general since the ``step size" for a fit is essentially only limited by machine precision (barring the effect of detector noise). For with and without detector noise effects, the lookup table RMSE value generally decreases as $\sigma$ increases for a fixed step size because there is more variation in intensity patterns over the pixels the larger $\sigma$ is, meaning the lookup table captures less of this breadth of variation and is thus less robust.  The variation in RMSE values may have some sampling (e.g., $\sigma$) dependence.  The RMSE value for the fitting algorithm decreases as $\sigma$ increases, with and without detector noise effects, as expected since the sampling gets better.

For the trials without detector noise (bottom part of table), the lookup table won only for the case of $\sigma=0.2$ pixels.  We have confirmed that for finer step size the lookup table wins for more of the values of $\sigma \leq 0.85$ pixels.  In the more realistic case of the inclusion of noise, though (top portion of table), the lookup table method with step size of 0.1 pixels clearly outperforms the fitting algorithm method for centroiding undersampled PSFs.  

For severely undersampled PSFs, one may be concerned with how one-to-one (injective) the mapping is from lookup table key (the simulated PSF with a certain centroid position) to value (the centroid coordinates) since the variation across pixels for a PSF lessens as $\sigma$ decreases.  If injectivity gets bad enough, the RMSE between the centroids found using the lookup table and the true centroid coordinates should suffer, but the results from Table \ref{noise_table} indicate good injectivity for PSFs with noise.  

For our examination of the injectivity of the lookup table, we generated 100 PSFs, each with a centroid located at a slightly shifted position with respect to a unique lookup table value, thus ensuring that the centroid found with the lookup table method should be unique if there is perfect injectivity.  We used a shift of 0.03 pixels, and we used simulated PSFs with no noise since we were testing the injectivity of the lookup table method, which is made with noiseless PSFs.  In Table \ref{injectivity}, we show the results of our examination of the injectivity of the lookup tables.  The lookup tables retain injectivity for PSFs below Nyquist sampling down to $\sigma = 0.2$ pixels, where
some injectivity is lost (i.e., the lookup table returns fewer than 100 unique centroids out of 100 PSFs used). Finer sampling step sizes (less than 0.1 pixels) increase the robustness of the lookup table only to an extent because it also decreases overall injectivity, so there is a limit to the improvement of a lookup table by finer sampling step size.  The one case which is not perfectly injective is still injective enough to provide good results, namely in the top portion of Table \ref{noise_table}.

\begin{table}
  \begin{center}
    \caption{Summary of the lookup table RMSE and the fitting algorithm RMSE results.  The sampling step size for the lookup table was 0.1 pixels, and 900 Gaussian PSFs were generated at random positions with randomized amplitude.}
  \label{noise_table}
     \begin{tabular}{l | p{2in} | p{2in}}
      \toprule 
      \Centering{$\sigma$ of Gaussians, in pixels}   & \Centering{RMSE $\pm$ Std Dev of RMSE, With Noise, Lookup Table} & \Centering{RMSE $\pm$ Std Dev of RMSE, With Noise, Fitting Algorithm} \\
      \hline
      \midrule 
       \Centering{0.2} &  \Centering{ 0.289 $\pm$ 0.007} &  \Centering{$0.488 \pm 0.012$}\\
      \Centering{0.4} &  \Centering{$0.0400 \pm 0.0009$} &   \Centering{$0.255 \pm 0.006$} \\
      \Centering{0.8} &  \Centering{$0.0273 \pm 0.0006$} &   \Centering{$0.0898 \pm 0.0021$} \\
      \Centering{2} &  \Centering{$0.0338 \pm 0.0008$} &   \Centering{$0.0183 \pm 0.0004$} \\
      \hline
	{$\sigma$ of Gaussians, in pixels}   & \Centering{RMSE $\pm$ Std Dev of RMSE, Without Noise, Lookup Table} & \Centering{RMSE $\pm$ Std Dev of RMSE, Without Noise, Fitting Algorithm} \\
	\hline
	\midrule
	 \Centering{0.2} &   \Centering{$0.0373 \pm 0.0009$} &  \Centering{0.0603 $\pm$ 0.0014} \\ 
       \Centering{0.4} &   \Centering{$0.0248 \pm 0.0006$} &  \Centering{0.0141 $\pm$ 0.0003} \\
       \Centering{0.8} &   \Centering{$0.0306 \pm 0.0007$} &  \Centering{0.000245 $\pm$ 0.000006} \\
       \Centering{2} &   \Centering{$0.0293 \pm 0.0007$} &  \Centering{0.000000129 $\pm$ 0.000000003} \\
    \end{tabular}
  \end{center}
\end{table}

\begin{table}
  \begin{center}
    \caption{Results of the one-to-one (injectivity) tests for the lookup table method.  The sampling step size for the lookup tables was 0.1 pixels, and 100 noiseless Gaussian PSFs were generated at positions slightly shifted from positions included in the lookup table so that 100 unique centroid positions should have resulted using the lookup table.  The lookup tables retain injectivity for PSFs below Nyquist sampling down to $\sigma=0.2$, where some injectivity is lost for this choice of sampling step size.}
  \label{injectivity}
     \begin{tabular}{l | p{2in} }
      \toprule 
      \Centering{$\sigma$ of Gaussians, in pixels}   & \Centering{Number of Unique Centroids out of 100}  \\
      \hline
      \midrule 
       \Centering{0.2} &  \Centering{ 87} \\
      \Centering{0.4} &  \Centering{100}\\
      \Centering{0.8} &  \Centering{100}\\
      \Centering{2} &  \Centering{100}\\
    \end{tabular}
  \end{center}
\end{table}

\section{CONCLUSION}
We have introduced the lookup table method of centroiding PSFs and illustrated its utility and accuracy over traditional fitting methods for undersampled PSFs.  This method is especially useful for cases when dithering is not desired or not possible.  The lookup table can be made even more accurate by decreasing the step size used to generate the lookup table, as long as the step size is not too small so that the lookup table does not lose too much injectivity.  

This work shows merely a simple example application of this method.  For a given observation scene, the expected PSF for a specific optical system can be modelled or obtained through examination with an interferometer, and a lookup table can be populated with entries of the form of this characteristic PSF at different non-integral centroid positions for different rotational orientations, for tip/tilt orientations of the detector, and for whatever idiosyncratic wavefront characteristics that may be expected. If the number of entries becomes unwieldy, multiple tables can be generated and checked during centroiding.  

For an observation scene with multiple PSFs, one can use a practical approach to centroiding.  One could locate a small sub-frame of an observation frame with the brightest pixel of a PSF in the center, and a script can use the lookup table at sub-frames shifted by 1 pixel in all directions to ensure the PSF is positioned the same way as in the lookup table keys.  The lookup table method described in this work can then be used on the sub-frame, and this process can be repeated for all the other PSFs.  We hope this method may one of many useful tools for accurate centroiding in different situations.

\section{Code, Data, and Materials Availability}
The script that does the analysis in this paper, {\tt lookup\textunderscore script.py}, is freely available here:  

{https://github.com/kevinludwick/lookup\_script.py} 

\noindent The script utilizes the module {\tt emccd\textunderscore detect}, which must be installed before using the script.  It is publicly available:

\noindent  {\tt emccd\textunderscore detect} freely available here:
 
{https://github.com/wfirst-cgi/emccd\_detect}

\noindent Simple installation instructions are included.  All code is in Python.


\bibliography{report} 

\bibliographystyle{spiebib} 

\end{document}